\documentclass[twocolumn]{aastex701}
\usepackage{graphicx} 
\usepackage{amsmath}
\usepackage{tikz}
\usetikzlibrary{shapes.geometric, arrows}
\usetikzlibrary{positioning}
\usetikzlibrary {arrows.meta}

\usepackage{hyperref}

\begin{document}

\title{The AGN Channel in 3D: Scattering Belts and the Importance of Eccentricity in the Black Hole Population}

\author[orcid=0000-0003-0384-018X]{Connar Rowan}
\affiliation{Niels Bohr International Academy, Niels Bohr Institute, Blegdamsvej 17, DK-2100 Copenhagen Ø, Denmark} 
\email[show]{connar.rowan@nbi.ku.dk}

\author[orcid=0000-0001-8716-3563]{Martin E. Pessah}
\affiliation{Niels Bohr International Academy, Niels Bohr Institute, Blegdamsvej 17, DK-2100 Copenhagen Ø, Denmark} 
\email[show]{mpessah@nbi.ku.dk}

\begin{abstract}
    Active galactic nuclei (AGN) are a promising origin for observed gravitational wave mergers. Current population synthesis models are limited to 1D N-body or Monte Carlo methods which rely on statistical approaches to resolving dynamical scatterings. We present three-dimensional hybrid $N$-body simulations of a population of black holes (BHs) surrounding an AGN using a new code in development, \texttt{AGNBI}, where close interactions are directly simulated for both single and binary objects. Our results show binary formations occur throughout the AGN disk, particularly where the migration time is long. While BHs efficiently migrate and pair up in migration traps, they are frequently disrupted by ionisation or exchanges from binary-single and binary-binary interactions. We report a broadened radial distribution of formations and mergers about the trap radius when accounting for binary-single and binary-binary interactions, which we describe as a ``scattering belt''. We assess the influence of the pre-existing BH population. When we model a non-zero eccentricity distribution in the BH population, the binary formations and mergers are suppressed by an order of magnitude compared to a non-eccentric population as fewer BHs embed in the AGN disk. We report an approximate merger rate of $\mathcal{R}_\mathrm{GW}\approx2\text{--}12\,\mathrm{Gpc}^{-3}\,\mathrm{yr}^{-1}$ with maximum BH masses of $10^2\text{--}10^3\,M_\odot$ through hierarchical mergers, where lower initial velocity dispersions produce larger BHs. The mass and mass ratio distributions are comparatively flat, suggesting AGN may be more relevant for high-mass or unequal-mass gravitational wave sources. 
\end{abstract}
\keywords{\uat{gravitational waves}{678} --- \uat{Galaxy accretion disks}{562} --- \uat{Active galactic nuclei}{16} --- \uat{High energy astrophysics}{739} --- \uat{N-body simulations}{1083} --- \uat{Black holes}{162}}
\section{Introduction}
The astrophysical origin of gravitational wave detections remains an open question, particularly for the extreme cases involving high chirp masses \citep{LIGO2020_GW190521_high_mass}, highly unequal masses \citep{Abbott2020_high_q} and high spins \citep{Zackay2019_high_spin,Abac2025_high_spin}. There are several proposed astrophysical origins: isolated field binary/triple evolution \citep{Antonini+2017}, nuclear \citep{Hopman2006,Antonini+2016,Mapelli2021} and globular \citep{DiCarlo2020,Antonini2020_globular_clusters,Mapelli2022_clusters} star clusters and active galactic nuclei \citep{OLeary+2009,Mckernan2012,Mckernan2014,McKernan2020,Bartos2017,Tagawa2020,Rowan2022,Rowan2024_rates,ford2022_AGNrates,Delfavero2024,Vaccaro2024,Vaccaro_2025_letter,Vaccaro2026_traps}. While the isolated channel is well positioned to explain the low-mass, low-spin mergers, the dynamical channels (AGN and star clusters) can more readily explain the abundance of high-mass mergers such as GW190521 and GW231123 \citep[][]{Abac2025_high_mass_merge} through successive mergers and in the case of AGN, the mass accretion of gas \citep[][]{Bartos2026_accretion_spin}. 

At present, the AGN channel is still poorly constrained, with reported merger rates spanning over five orders of magnitude \citep[see review in][]{Mandel2022_rates_review}. This significant uncertainty largely results from the environment's complexity. The system comprises several components (the supermassive black hole (SMBH) and its accretion disk, a black hole (BH) population, a binary black hole (BBH) population and a stellar population). These components interact with each other through a wide array of physical processes (e.g., single-single/binary-single/binary-binary interactions, gas accretion, Kozai-Lidov oscillations \citep[e.g.,][]{Smadar2016}, gas-assisted binary formation, out-of-plane interactions with the AGN disk via transits, in-plane migration, disk-embedded binary evolution and migration traps among many others). The properties of the AGN disk models, which govern how these processes play out, vary considerably depending on the assumptions made (e.g., \citealt{Sirko2003} and \citealt{Thompson_disc_2005}) and their initial parameters \citep[e.g.,][]{Gangardt2024,Vaccaro2024,Mckernan2025_mcfacts}. 

The current state-of-the-art approaches to modelling the BH population of an AGN and the resulting mergers can be divided into three categories: 1) \textit{semi-analytical} \citep[e.g.,][]{Bartos2017, Yang2019_lett,Vaccaro_2025_letter, Vaccaro2024} 2) \textit{Monte Carlo methods} \citep[e.g.,][]{Yang2019,McKernan2020_AGNmontecarlo,McKernan2020,Rowan2024_rates,Xue2025} and 3) \textit{1D N-body simulations} \citep[e.g.,][]{Tagawa2020, Tagawa2020_spin, Tagawa+2021_eccentricity, Mckernan2025_mcfacts, Cook2025_mcfacts, Delfavero2024}.

We present a new approach to modelling the AGN channel population, using a new purpose-built code \textit{Active Galactic Nucleus Body Integrator}: 
\texttt{AGNBI}. This code utilises an $N$-body framework designed to simulate the dynamical and orbital evolution of stellar-mass black holes (BHs) embedded within an AGN accretion disk on sub-orbital timescales in 3D. The code tracks a population of single and binary black holes (BBHs), evolving their Keplerian elements subject to gas-dynamical interactions, single-single/binary-single/binary-binary encounters (hereafter SS, BS and BB respectively), in addition to tracking the inner orbital evolution of BBHs via gaseous and gravitational wave dissipation. Crucially, we accurately model the meaningful inter-object scattering interactions by simulating them directly. 

The ethos of the approach taken here is not to capture \textit{all} the encounters, but rather to isolate those which dominate the formation and destruction of binaries in the AGN disk. This approach allows us, for the first time, to analyse how a full AGN population of BHs approach and scatter off each other in 3D, including at migration traps.

\section{Methods}
\label{sec:methods}
\subsection{Fiducial AGN disk and BH population}
\label{sec:initial_conditions}

We consider an AGN with a supermassive black hole (SMBH) of mass $M_\bullet=4\times10^{6}\,M_\odot$ hosting a population of 20,000 BHs \citep{Miralda2000} for an assumed AGN lifetime of $t_\mathrm{AGN}=10^7\,\mathrm{yr}$. The fiducial model is initialized by populating a predefined AGN disk model with the stellar-mass black hole population with masses $m$ according to $N(m) \propto m^{-2}$ with $m\in[5\,M_\odot,40\,M_\odot]$ and semi-major axes $a$ according to $N(a) \propto a^{-0.5}$ with $a\in[10^3r_\mathrm{g},10^7r_\mathrm{g}]$, where $r_\mathrm{g}=2GM_\bullet/c^2$ is the Schwarzschild radius of the SMBH, $c$ is the speed of light and $G$ is the gravitational constant. For our fiducial model, in line with \cite{Tagawa2020} and \cite{Xue2025}, we sample the inclination ($i$) assuming a Gaussian velocity dispersion of $\sigma_i=\beta/\sqrt{3}$
\begin{equation}
    \frac{dN}{d\sin(i)} = \frac{1}{\sqrt{2\pi}\sigma_i}\exp\bigg[-\frac{\sin^2(i)}{2\sigma_i^2}\bigg]\,,
\end{equation}
where $\beta\in[0,1]$ controls the initial dispersion, taking $\beta=1$ for our fiducial case. Now in 3D, we relax the assumption that the orbital eccentricities of BHs and BBHs about the SMBH are zero and assume the system is relaxed ($\sigma_e\approx2\sigma_i$) and the eccentricity follows a Rayleigh distribution \citep{Trani2025_turbulence} with $f(e)=e/\sigma_e^2  \exp[-e^2/(2\sigma_e^2)]$, which naturally transitions to a thermal-like distribution when $\sigma_\mathrm{e}$ is large. Since this distribution allows for eccentricities greater than one, we reject and re-sample such cases. The orbital angles---the longitude of the ascending node, argument of periapsis, and mean anomaly ($M_0$)---are drawn uniformly from $[0, 2\pi)$. In order to more clearly assess how binaries form and merge through the gas-capture formation mechanism, we assume no pre-existing binaries in this work.

\subsection{Orbit propagation}
The code employs a hybrid integration scheme. The global motion of singles and binaries around the SMBH is treated analytically using Keplerian elements, updated incrementally by secular and impulsive source terms. 

Using a global timestep, the mean anomaly is updated via $M = M_0 + n(t - t_0)$, where $n = \sqrt{GM_{\text{enc}} / a^3}$ is the mean motion and $M_{\text{enc}}$ is the enclosed mass (incorporating the disk mass and mass of the cluster\footnote{We compute the enclosed disk and population mass by integrating over the radial density profile of the disk and the object population up to the semi-major axis of an object in the simulation. For the disk, we also compute the quadrupole moment of the disk and compute the orbit-averaged precession rate on each object.}). The eccentric anomaly $E$ is solved via a Newton-Raphson iteration of Kepler's equation, $M = E - e \sin E$, to map the orbital elements to Cartesian phase space when required by the collision handler (Section~\ref{sec:collisions}). By propagating the orbits using the orbital elements we simultaneously avoid the long-term energy and angular momentum drifts associated with directly integrating orbits under the SMBH potential and reduce the computational expense.

\subsection{Gas-Dynamical Source Terms}
The evolution of the orbital elements $(a, e, i, \Omega, \omega)$ is governed by the local disk properties—density ($\rho$), surface density ($\Sigma$), scale height ($H$), and sound speed ($c_s$)—interpolated from a 1D lookup table in radius $R$, generated using the Python package \texttt{pagn} \citep[][]{Gangardt2024}. The disk equations are solved according to a fiducial model where we assume a viscosity parameter $\alpha=0.1$, the SMBH has an Eddington accretion fraction of 0.1 and radiative efficiency of 0.1. 
\subsubsection{Disk Transits}
For BHs and BBHs on inclined orbits with $\sin(i) > H(R)/a$, aerodynamic drag is imparted impulsively during disk transits. The vector velocity perturbation $\Delta \mathbf{v}$ during a transit of duration $t_{\text{cross}}$ is computed as:
\begin{equation}
    \Delta \mathbf{v} = - \frac{\dot{m}}{m} \mathbf{v}_{\text{rel}} t_{\text{cross}}\,,
    \label{eq:impulse}
\end{equation}
with the accretion rate $\dot{m}= \pi R_\mathrm{acc}^2\rho v_\mathrm{eff}$, consistent with supersonic motion.
The effective accretion radius is the Bondi-Hoyle-Lyttelton radius, truncated by the Hill sphere: $R_{\text{acc}} = \min(r_{\text{BHL}}, r_{\text{H}})$ with
\begin{equation}
    r_\mathrm{BHL} = \frac{2 G m}{v_\mathrm{eff}^2}, \quad r_\mathrm{H} = a\bigg(\frac{m}{3M_\bullet}\bigg)^{1/3}\,,
\end{equation}
and $v_\mathrm{eff} = \sqrt{v_\mathrm{rel}^2+c_\mathrm{s}^2}$.

We compute $\Delta \mathbf{v}$ as a vector individually for both disk crossing points, accounting for $a$, $e$ and $i$, where the crossing time $ t_\mathrm{cross}=(2H)/|v_{z,0}|$, $v_{z,0}$ is the midplane vertical velocity at the crossing and $\boldsymbol{v}_\mathrm{rel}$ is the difference between the velocity vector of the BH/BBH and the Keplerian orbit of the gas at its intersection of the disk. The perturbation from both crossings is mapped back to changes in the Keplerian elements $(\Delta a, \Delta e, \Delta i, \Delta \Omega, \Delta \omega)$, which are subsequently orbit-averaged\footnote{If an orbit is significantly modified by a single crossing then this would invalidate our orbit-averaging assumption, however we find no instance of this in our simulations before an object transitions to the migration regime, once embedded.}.

\subsubsection{Orbital Migration and Damping}
For fully embedded BHs and BBHs,  $\sin(i) <H(R)/a$, we apply the migration torques of \cite{Jiminez_2017_torques} and thermal torques of \cite{Grishin2024}. The secular change in the semi-major axis is driven by the total torque $\Gamma_{\text{tot}}$:
\begin{equation}
    \dot{a} = \frac{2}{m}\sqrt{\frac{a}{G M_{\text{enc}}}} \Gamma_{\text{tot}}\,.
\end{equation}
The inclination and eccentricity damping timescales ($\tau_i$ and $\tau_e$ respectively) scale with the migration timescale $\tau_\mathrm{mig}$ and the disk's aspect ratio \citep{Kanagawa2020}:
\begin{equation}
    \tau_e = \frac{|\tau_{\rm mig}|}{0.780} \left( \frac{H}{a} \right)^2\,,
    \label{eq:i_e_damp}
\end{equation}

\begin{equation}
    \tau_i = \frac{|\tau_{\rm mig}|}{0.544} \left( \frac{H}{a} \right)^2\,.
    \label{eq:i_damp}
\end{equation}
with
\begin{equation}
    \tau_{\rm mig} = \frac{m\sqrt{G M_\mathrm{enc}a}}{2|\Gamma_{\rm tot}|}\,.
\end{equation}
The inclination and eccentricity are then damped at a continuous rate: $\text{d}e/\text{d}t= - e/\tau_\mathrm{e}$, $\text{d}i/\text{d}t=-i/\tau_\mathrm{i}$.
Mass accretion onto the BHs is given by the BHL rate, bounded by the Eddington limit:
\begin{equation}
    \dot{m} = \min \left( \dot{m}_\mathrm{BHL}, \dot{M}_{\text{Edd}} \right)\,.
\end{equation}
For transiting objects, their accretion is calculated through the same impulse approach as in Eq.~\eqref{eq:impulse}.
\subsection{Binary Black Hole Evolution}
The inner binary semi-major axes of the BBHs ($a_{\text{bin}}$) are evolved subject to the gaseous torques of their circumbinary disks, as proposed for embedded binaries by \cite{Ishibashi2020}:
\begin{equation}
    \frac{da_\mathrm{bin}}{dt}\bigg|_\mathrm{gas}= -\frac{24\pi\alpha c_\mathrm{s}^{2}\Sigma(1+e_\mathrm{bin})^2}{\mu \Omega_\mathrm{bin}}a_\mathrm{bin}\,,
    \label{eq:a_bin}
\end{equation}
where $m_\mathrm{bin}=m_1+m_2$, $\mu=m_1m_2/m_\mathrm{bin}$ is the reduced mass, and $\Omega_\mathrm{bin}=\sqrt{Gm_\mathrm{bin}/a_\mathrm{bin}^3}$. We assume the inner orbital eccentricity $e_\mathrm{bin}=0$. The semi-major axis of a BBH, $a_\mathrm{bin}$, also experiences shrinkage due to gravitational wave dissipation according to \cite{Peters1964}, where we sub-cycle the binary evolution to maintain the integration accuracy. 

We assume a binary merges when $a_\mathrm{bin}<6Gm_\mathrm{bin}/c^2$ and the resulting new single object retains the mass of the binary. We apply a simple velocity kick according to \cite{Gonzalez2007} in a random direction in 3D with
\begin{equation}
    v_\mathrm{kick}=1.2\times 10^4\, Q^2\sqrt{1-4Q}(1-0.93Q)\, \mathrm{km}\,\mathrm{s}^{-1}\,,
\end{equation}
where $Q=q/(1+q)^2$, $q=m_2/m_1$ is the mass ratio of the binary ($m_2\leq m_1$) and we enforce a minimum kick of $10\,\mathrm{km}\,\mathrm{s}^{-1}$. The inner binary inclination $i_\mathrm{bin}$ (set by their formation, see Section~\ref{sec:collisions}) is evolved according to the calculations in \cite{Fabj2026_realignment}. We assume their fiducial realignment efficiency.
\subsection{Encounters and Collisions}
\label{sec:collisions}
To maintain $O(N)$ computational scaling, the code uses a log-spherical spatial hashing algorithm. Space is discretized into bins in radius ($r$), azimuth ($\phi$) and angle above the midplane ($\theta$). Objects occupying the same or adjacent bins are flagged as candidate encounters. An encounter is triggered if the separation between two objects is less than the scattering kernel of $R_\mathrm{scat}=3r_{\text{H}}$. Because objects are advanced along Keplerian tracks, faster encounters may significantly overstep the collision threshold. We mitigate this by backtracking the objects to exactly when they crossed the kernel. Once synchronized, the localised encounter is integrated in 3D Cartesian space using a 4th-order Runge-Kutta integrator with an adaptive timestep given by $dt=0.015 \times\min_{i<j}(\Delta r_{ij}/\Delta v_{ij})$ where the $\Delta r_{ij}$ and $\Delta v_{ij}$ are the separations and relative velocities between each object pair in the few-body simulation. Crucially, if the encounter occurs within the disk, the integration continuously evaluates the gas drag on each body using the background Keplerian gas velocity field according to \cite{Ostriker1999}. 

For SS encounters, the sub-grid integration terminates when the bodies exit the Hill sphere or become permanently gravitationally bound, which we define as\footnote{When assuming a softer binding criterion of $a_\mathrm{bound}<0.5 r_\mathrm{H}$ we form binaries that would otherwise decouple if we set $a_\mathrm{bound}<0.25 r_\mathrm{H}$ if the orbit is eccentric. So we maintain this stricter requirement.} $a_\mathrm{bound}<0.25 r_\mathrm{H}$. Bound systems are converted into new binary objects, capturing dynamical formation channels within the AGN disk. For BS or BB encounters, we treat binaries as single objects unless the separation $\Delta r$ between any object and a binary is $\Delta r < 5 a_\mathrm{bin}$, in which case we split the binary into its components, selecting a random phase, ascending node and argument of periapsis. In this way we capture exchanges, ionisations and perturbations to the binary elements directly. In all cases, we record when a BH/BBH becomes unbound from the other objects and reinstate it into the global simulation. 
A limitation of our kernel-based approach to collisions is that we do not accurately capture the effect of potential mean-motion resonances that occur between satellites in the AGN disk \citep[e.g.,][]{Moncrieff2026}, although the stability of these resonances when accounting for the many objects is not guaranteed \citep[see][]{Epstein-Martin2025, Gonglewski2026}.
\begin{figure*}
\centering
       \includegraphics[width=0.94\textwidth]{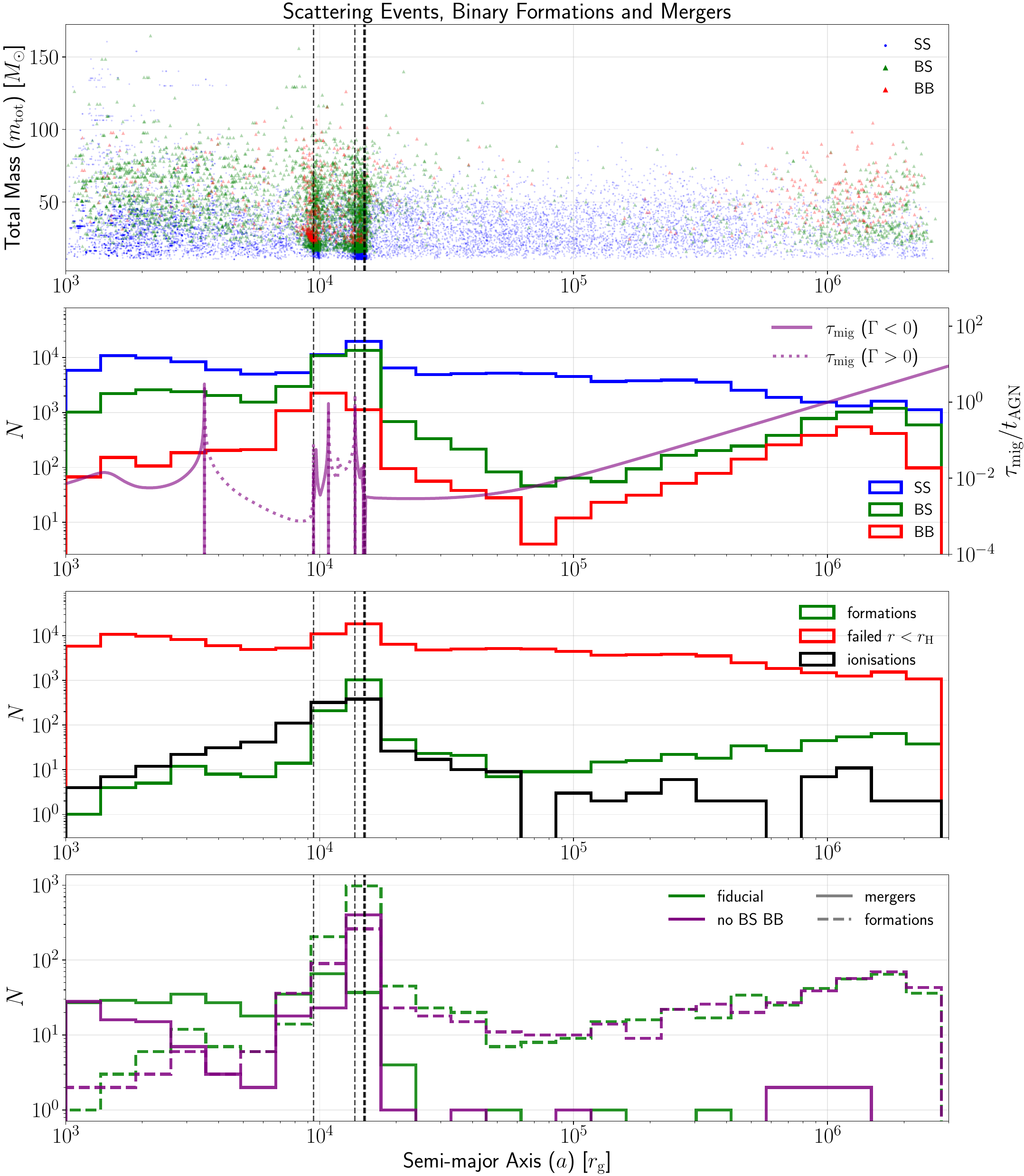} 
    \caption{Distribution of scattering events and mergers in semi-major axis $a$. The vertical dashed lines show the location of migration traps according to the disk model. \textit{1st row:} The radial position and total mass of single-single (SS), binary-single (BS) and binary-binary (BB) encounters. We show only one in ten encounters of each type (sampled at random) for visual clarity.
    \textit{2nd row:} The distribution of SS, BS and BB encounters in bins of $a$ and the migration time (purple) for objects embedded in the disk. \textit{3rd row:} Semi-major axis of binary formations (green), ionisations (black) and encounters that reach a separation $<r_\mathrm{H}$ but fail to form binaries (red). \textit{4th row:}  The radial position of mergers and formations for our fiducial run (red) and a separate run where BS and BB encounters are switched off (green).}
    \label{fig:scatterings}
\end{figure*}
\subsection{Encounter Velocity Threshold}
\label{sec:threshold}

The global timestep is enforced by the need to temporally resolve the crossings of the scattering kernel. To maintain computational efficiency while capturing the highly non-linear dynamics (especially relevant for BS and BB interactions), we impose a strict upper limit on the relative velocity of encounters processed by the few-body integrator. Encounters are rejected prior to integration if their initial relative velocity exceeds a threshold of $v_{\text{rel}} > 10 v_{\text{H}}$, where $v_\mathrm{H}=\sqrt{Gm_\mathrm{bin}/r_\mathrm{H}}$. For SS encounters, this criterion encapsulates all binary-formation events and strongly gravitationally focused encounters, while excluding encounters in the purely impulsive regime.

Considering binary encounters, the minimum internal orbital velocity of a marginally stable binary (e.g., $a_{\text{bin}} \approx 0.25 r_{\text{H}}$) is related to the local Hill velocity:
\begin{equation}
    v_{\text{bin}}^2 \approx \frac{G m_{\text{bin}}}{0.25r_{\text{H}}} \approx 4 v_{\text{H}}^2\,.
\end{equation}
An incoming single at our maximum velocity boundary possesses a specific kinetic energy of $\sim 100 v_{\text{H}}^2$. While this is energetic, it sets the binding energy at which the binary is effectively considered hard at $|E_{\text{hard}}| > 100 v_{\text{H}}^2$, which is still over an order of magnitude below the mean energy of the cluster $E_\mathrm{\sigma}$:
\begin{equation}
 \frac{E_\mathrm{\sigma}}{E_\mathrm{H}} \approx \frac{\sigma^2}{v^2_\mathrm{H}}=\frac{\beta^2}{9}\bigg(\frac{3M_\bullet}{m_\mathrm{bin}}\bigg)^{2/3}\approx1200,
\end{equation}
where we have assumed $\beta=1$ and $m_\mathrm{bin}=10\,M_\odot.$

In short, our choice to truncate the relative velocities of encounters at $10 v_{\text{H}}$ ensures that all non-linear BS and BB interactions (exchanges, hardening, and chaotic resonances) are captured, however we do not consider encounters with energies $E>100v_\mathrm{H}^2$ and thus binaries are typically not ionised beyond an equivalent binding energy. 
\subsection{Timestepping}
To safely bracket and resolve the critical interactions within the Hill sphere, the global baseline timestep is defined as:
\begin{equation}
    \Delta t_{0} = \xi\frac{R_\mathrm{scat}}{v_\mathrm{enc}}=\xi \frac{3r_\mathrm{H}}{\frac{3}{2}r_\mathrm{H}\Omega(a_\mathrm{min})}=\xi\frac{2}{\Omega(a_\mathrm{min})}\,,
    \label{eq:global_timestep}
\end{equation}
where $\Omega=\sqrt{GM_\mathrm{enc}/a^3}$ and $a_\mathrm{min}$ is the inner boundary (here $10^3r_\mathrm{g}$). A timestep with $\xi<1.0$ would capture all interactions between objects on circular coplanar orbits. We set $\xi=0.5$, giving $\Delta t_0 \approx 0.06\,\mathrm{yr}$, which detects encounters with $\sqrt{i^2+e^2}<0.05$ at the inner boundary and \textit{all} encounters from $a\gtrsim 10^{4}r_\mathrm{g}$, which crucially includes the migration traps (see below). For our fiducial model, this results in $\sim10^{6}\text{--}10^{7}$ scatterings, which we examine in detail below.
The source terms, which evolve the orbits only secularly, utilise an adaptive timestep based on the local orbital period, $\Delta t_i = \Delta t_{\text{0}} (a_i / a_{\text{min}})^{1.5}$, limiting their computational expense. We sub-cycle the evolution of $a_\mathrm{bin}$ where necessary to accurately integrate Eq.~\eqref{eq:a_bin}. 

Since the inner orbital elements of our binaries evolve in time, we naturally capture the initial conditions during our few-body integrations, where only the phases of binaries in binary-single or binary-binary interactions are arbitrarily chosen. Given a formed binary will undergo very many orbits before another encounter, the assumption of a random phase should not impact our results.
\section{Results}
\subsection{Scattering belts}
In many studies, migration traps play a central role in binary formation, with a significant fraction (sometimes the majority) of binaries forming near the trap radius because convergent migration enhances the local number density of objects there.
In 1D studies, this represents a delta function of objects at the zero-torque radius, where objects that reach this position are guaranteed to form a binary with any other object residing there \citep{Mckernan2025_mcfacts,Vaccaro2026_traps}, possibly leading to runaway mergers. However, this assumption neglects several key physical aspects of the problem, each of which serves to increase the time required for a binary to form as an object approaches another in the trap. 

Singles or formed binaries in the vicinity of a trap will scatter with other converging objects, potentially disrupting the radial flow of objects arriving at the trap and decreasing the likelihood of binary formation. Another aspect neglected in a 1D treatment is the ``catch-up'' time required for objects at different orbital phases to meet each other. If a binary fails to form, which is increasingly likely in lower-density environments \citep[e.g.,][]{Rowan2022,Whitehead2025_adiabatic}, then the individual objects must wait another libration time in order to try again (assuming they do not subsequently migrate away from each other in the disk).

In Figure~\ref{fig:scatterings} we show the semi-major axis distribution of SS, BS and BB scatterings and the location of BBH formations and mergers. We find a noticeable increase in scatterings at the migration trap in the disk, where $\sim10^{4}$ scatterings take place during the lifetime of the AGN. While SS encounters naturally dominate most regions of the disk, BS reach near parity with SS encounters at the migration trap, with BB encounters also within an order of magnitude. The number of SS scatterings naturally decreases moving away from the trap to higher semi-major axes. The distributions of BS and BB encounters, however, exhibit a smooth drop in the midsection of the disk at $10^5r_\mathrm{g}$. This results from the short migration time in this region of the disk, where the residence time of BBHs limits the number of BS and BB encounters. Beyond $10^5r_\mathrm{g}$, the migration time rises, becoming comparable to the AGN lifetime and facilitating more encounters between BBHs and other objects.
\begin{figure}[!htbp]
    \centering
    \includegraphics[width=\linewidth]{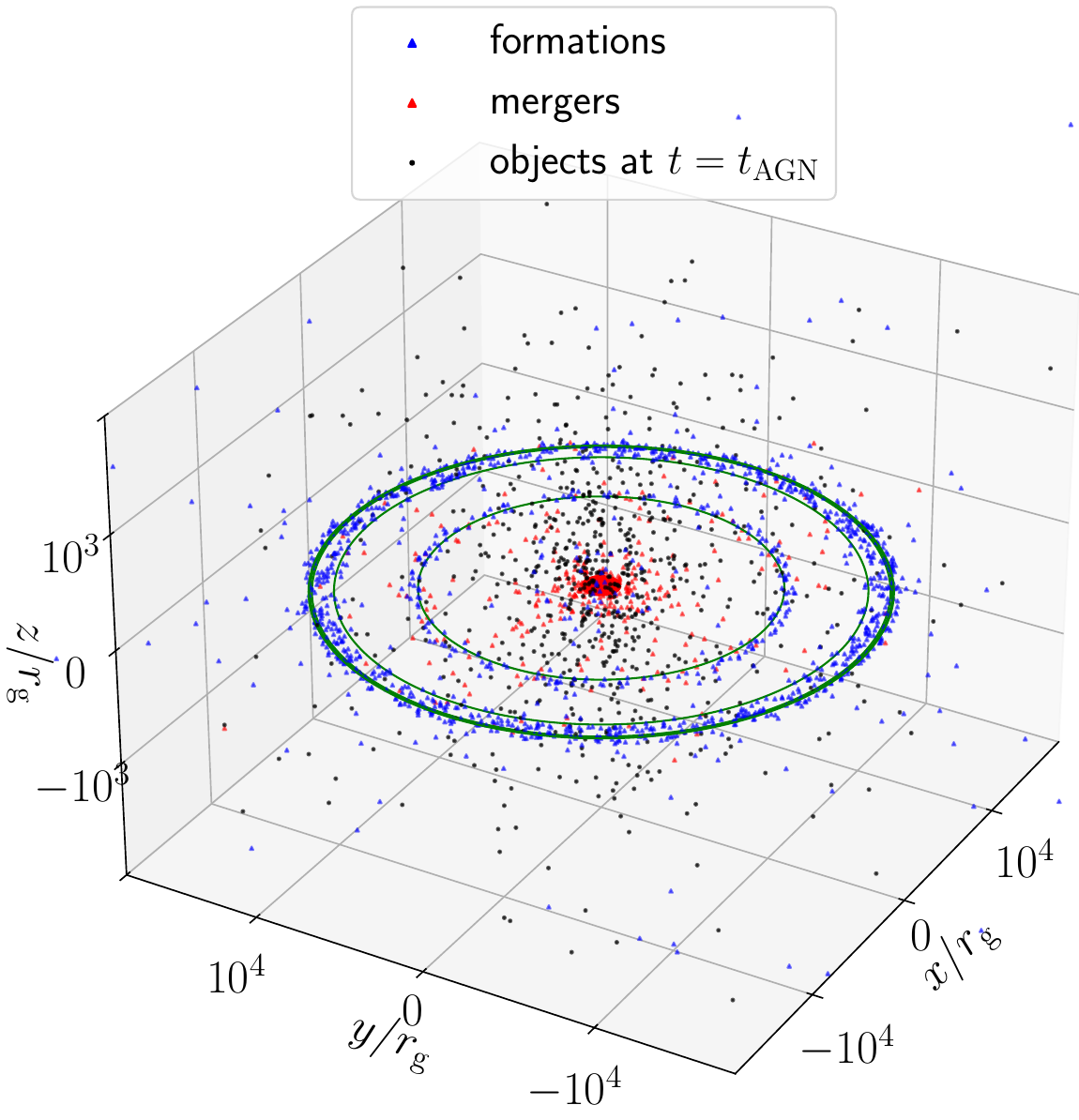}
    \caption{Visualisation of interactions at a migration trap in 3D. The 3D locations of BHs and BBHs at $t=t_\mathrm{AGN}$ are shown in black. The locations of formations across the full AGN lifetime are shown in blue and the locations of mergers in red. The stable migration traps are represented by the green rings for a $10\,M_\odot$ object. The central red dot represents our inner boundary condition.}
    \label{fig:3D_trap}
\end{figure}
The spike in interactions at the migration trap is coincident with a spike in BBH formation and ionisation. In fact, we find that $\sim 80\%$ of all formed binaries at the trap are subsequently destroyed through ionisation or exchanges, with many objects forming binaries that are later destroyed multiple times before merging. This fraction is likely highly dependent on the location of the trap, which can change dramatically with the value of $M_\bullet$ and $\alpha$. We reiterate that we detect all encounters at the migration trap, but those responsible for binary destruction are typically low-velocity ($\Delta v< 5\sqrt{G(m_i+m_j)/r_\mathrm{H}}$) encounters with another black hole or binary scatterings via exchanges or ionisation. We compare our fiducial run to another identical run where we switch off BS and BB encounters (Figure~\ref{fig:scatterings}, fourth panel). When we ignore both types of encounter, we observe a doubling in the merger rate and a much narrower spike at the migration trap. The number of formations is reduced without scatterings as each formed BBH cannot be ionised and form another BBH until it has merged. The less prominent peak in the merger distribution in our fiducial model stems from objects being scattered away from the trap and/or being ionised. We coin the term ``scattering belts'' to describe these traps as objects are typically unable to maintain their position long at the precise trap radius before they are scattered. The frequency of scatterings also leads to $\sim 10\%$ of our mergers occurring while they are in the transiting regime, i.e., outside of the disk. Depending on the mass of gas retained in these out-of-disk mergers, possible electromagnetic counterparts \citep[][]{Cabrera2026_AGN_EM} could have different properties to fully embedded mergers. These include a much shorter breakout time, as photons do not have to escape the dense AGN disk and stronger X-ray/UV emission if the retained gas parcel is sufficiently diffuse.
\begin{figure*}
\centering
\includegraphics[width=1.0\textwidth]{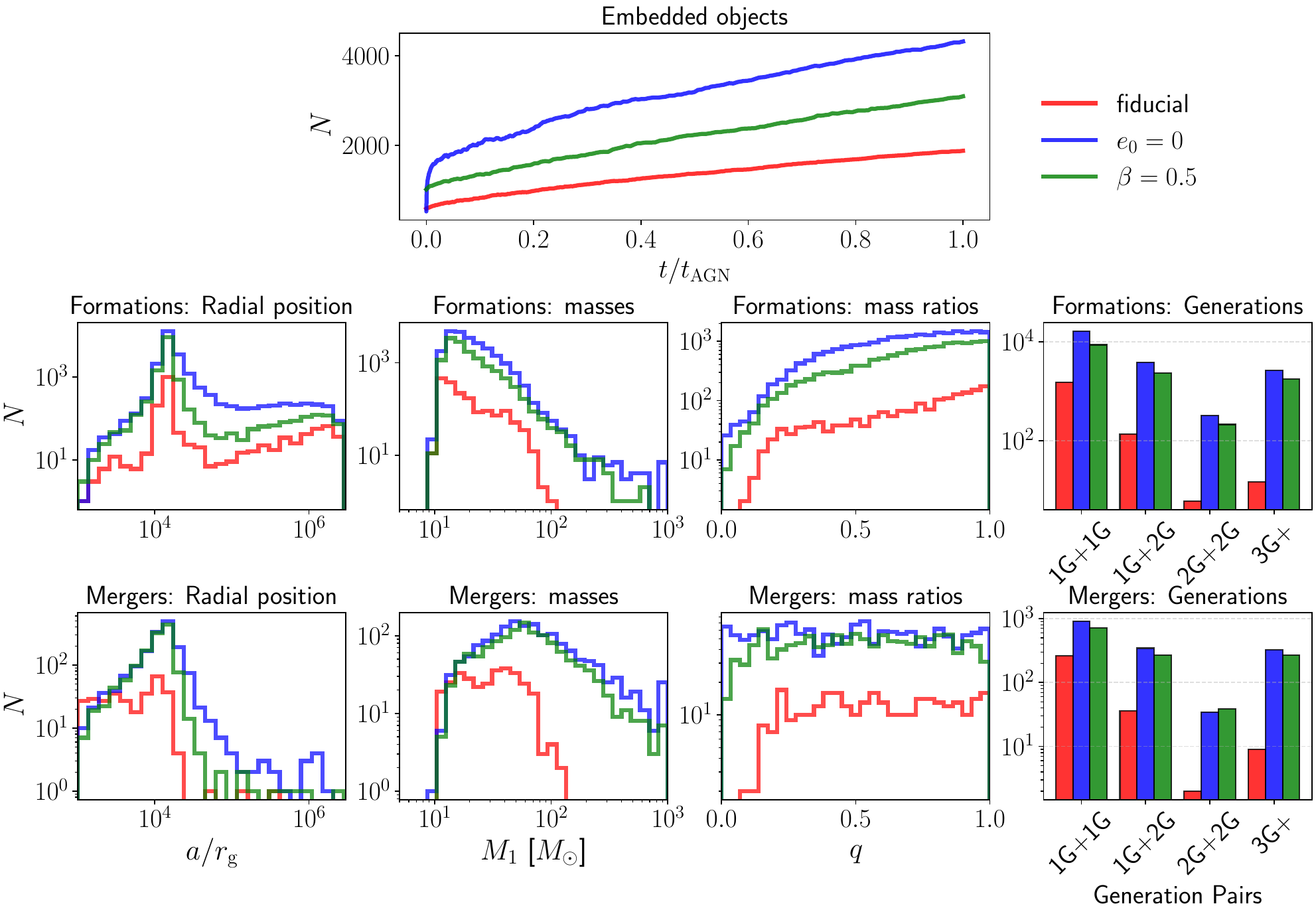} 
    \caption{Semi-major axis $a$, primary mass $m_\mathrm{1}$, mass ratio $q$ and merger generation for formations and mergers in our simulations: fiducial $\beta=1$ (red), $e_0=0$ (blue) and $\beta=0.5$ (green). In the top panel we show the number of objects embedded in the AGN disk ($\sin(i) < H(R)/a$) as a function of time.}
    \label{fig:eccentricity}
\end{figure*}

To better visualise the trap, we show the location of objects at $t=t_\mathrm{AGN}$ alongside the location of all mergers in 3D in Figure~\ref{fig:3D_trap}. As shown in the figure, the majority of formations in this volume of the simulation domain occur at or around the traps. The mergers and general population of objects are often found outside of the traps themselves due to scatterings. Due to the thin geometry of the disk, scattered objects damp their eccentricities and inclinations faster than their radial migration timescale (Eqs.~\eqref{eq:i_e_damp} and~\eqref{eq:i_damp}). Scattered objects therefore still have a low velocity dispersion in the vicinity of a trap as they migrate back towards it or (depending on the object's mass) towards the next trap.

\subsection{The role of BH population eccentricity}
The primary factor governing merger rates in the AGN channel is the size of the embedded BH population. The geometry of the disk and the phase space of the pre-existing BHs in the nuclear star cluster (prior to an AGN episode) dictate how many BHs embed within $t_\mathrm{AGN}$ and thereby influence the formation/merger efficiency. An often-neglected aspect of this phase space is the eccentricity of the initial BH population. The recent semi-analytical study of \cite{spieksma2025} finds that both the semi-major axis and eccentricity of a transiting BH follow a non-trivial evolution that is ultimately tied to its initial conditions. To our knowledge, the Monte Carlo simulations of \cite{Yang2019} represent the only population study of the AGN channel at present which accounts for eccentricity in the alignment time. They report the upper mass range of mergers in the region of $\sim150\,M_\odot$ under the assumption that any black hole that aligns automatically merges randomly with another that aligns from their sample at least once. In Figure~\ref{fig:eccentricity}, we compare our fiducial run to an identical run where we assume only circular initial orbits ($e_0=0$) and another with a colder population with $\beta=0.5$ where we again maintain $\sigma_e = 2\sigma_i$.

Our results indicate that eccentricity plays a very significant role in the efficiency of the AGN channel. The disparity between our fiducial and $e_0=0$ runs is significant, with an order of magnitude difference in the number of formations and mergers across the simulation domain, particularly for high generation numbers (3G+). The greater merger number also leads to the merger of much larger BBHs, with our fiducial run reporting a maximum binary mass of $\sim10^2\,M_\odot$ and $e_0=0$ run in the intermediate-mass black hole range of $\sim10^3\,M_\odot$. We note that our $e_0=0$ model with a wider inclination distribution ($\beta=1$) still embeds more objects than our colder $\beta=0.5$ model. Our $\beta=0.5$ model still indicates that the formation of low-mass intermediate-mass black holes is possible even when eccentricity is accounted for. We therefore stress the need to account for the eccentricity distribution of the initial black hole population in order to accurately model properties of the AGN channel.
\subsection{Influence of high-energy collisions on results}
Here, we test the survival of our merging binaries against the impulsive high-velocity encounters not captured in our simulation. Physically, these correspond to encounters between typically embedded binaries on near-circular orbits with still highly misaligned singles. As our criterion rejects encounters with relative velocities greater than $10v_\mathrm{H}$ we can be certain that all unsimulated encounters are in the impulsive limit (see Section~\ref{sec:threshold}). 

\begin{figure}[!htbp]
    \centering
    \includegraphics[width=\linewidth]{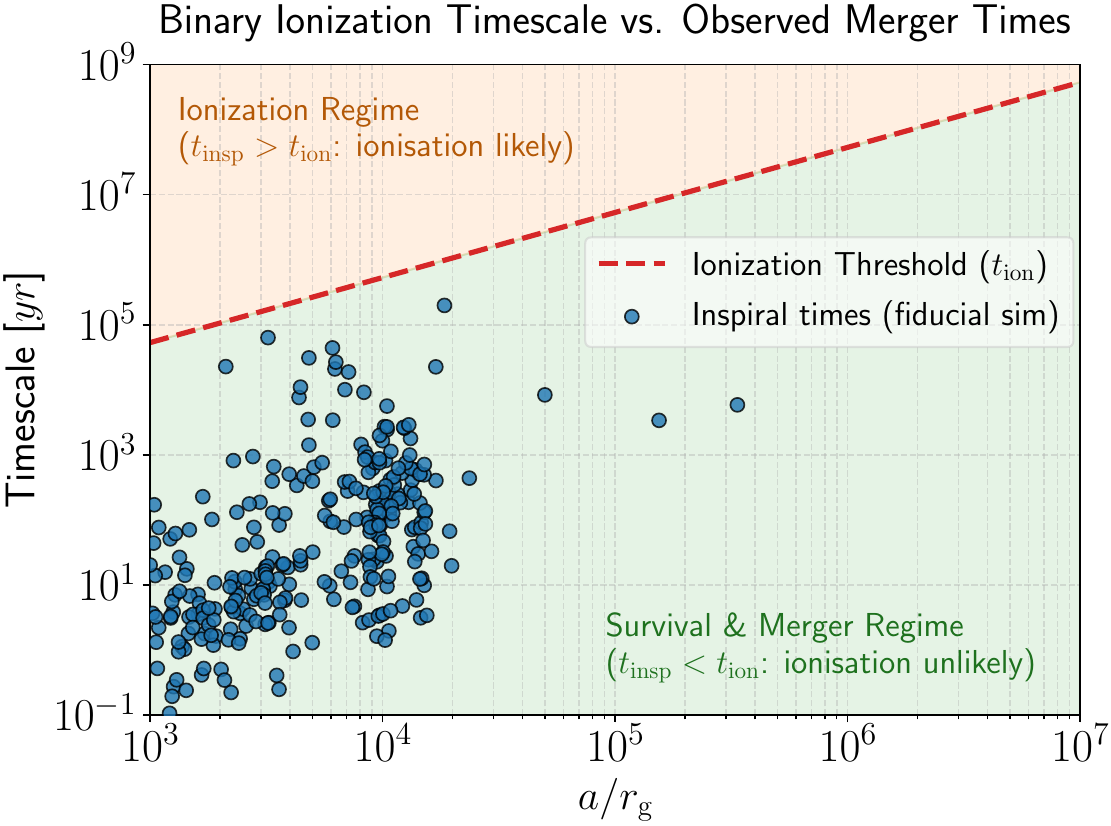}
    \caption{Comparison of the binary lifetimes in our simulation (blue points) and the typical ionisation timescale (Eq.~\eqref{eq:ion}), shown as the red dashed line. We have assumed that the binary and perturber are all $5\,M_\odot$ and the binary does not further harden beyond $0.25r_\mathrm{H}$.}
    \label{fig:ionisation}
\end{figure}

We follow the calculations of \cite{Hamilton2024} to compute the timescale required for a binary to undergo a deep encounter with an impact parameter smaller than the binary semi-major axis ($b<a_\mathrm{bin}$). Since the alignment time scales extremely strongly with the inclination $\sim \sin(i/2)^{4}$ \citep{Rowan2024_rates}, the portion of the cluster which does not align with the disk within $t_\mathrm{AGN}$ ($\sim 90\%$ of the cluster for our fiducial model) can be assumed to be effectively static with no modification from the AGN disk. Hence the minimum impact parameter for an encounter can be calculated as
\begin{equation}
    b_\mathrm{min}\sim\frac{1}{\sqrt{\pi n \sigma t_\mathrm{bin}}}\,,
\end{equation}
where $\sigma=(\beta/\sqrt{3})\sqrt{GM_\mathrm{enc}/a}$ is the velocity dispersion, $n$ is the midplane number density of objects for our Gaussian distribution of $\sin(i)$ and $t_\mathrm{bin}$ is the lifetime of the binary.
For a close encounter to unbind a binary, the impulsive kick must be on the order of the binary's inner orbital velocity, which occurs at
\begin{equation}
    b_\mathrm{ion}\sim\frac{2G^{1/2}m_3a_\mathrm{bin}^{1/2}}{m_\mathrm{bin}^{1/2}\sigma}\,,
\end{equation}
where $m_3$ is the mass of the incoming BH. We can equate $b_\mathrm{min}\sim b_\mathrm{ion}$ to solve for the typical timescale required for a formed binary to undergo an ionisation from the high-velocity population 
\begin{equation}
    t_{\mathrm{ion}} = \frac{m_\mathrm{bin} \sigma}{4 \pi G m_3^2 n a_{\mathrm{bin}}}\,.
    \label{eq:ion}
\end{equation}
In Figure~\ref{fig:ionisation} we show how the ionisation timescale $t_\mathrm{ion}$ varies with distance from the SMBH and overplot the time between formation and merger for the binaries in our fiducial simulation. We have assumed the binary does not harden beyond its initial formation at $a_\mathrm{bin}=0.25r_\mathrm{H}$, which would only increase the time required for a successful ionisation. 
From the figure, the typical ionisation timescale from these objects is longer than the lifetime of all of our merging binaries for the radial domain of our simulations. On this basis, we conclude that ignoring the high-velocity population will not meaningfully affect our results. 
For a more massive AGN system with $M_\bullet=10^8\,M_\odot$, where the stellar and black-hole number densities are substantially higher, this approximation may be less well justified. Note that this is only the influence of the black hole population, we do not consider the background stellar population which will also contribute to ionisations and potentially limit the merger rate.

\subsection{Overall properties of the AGN channel}
Considering the results shown in Figure~\ref{fig:eccentricity}, the merging mass and mass ratio distributions are flatter than those of isolated binary evolution, consistent with previous expectations \citep[e.g.,][]{Yang2019, Tagawa2020, Rowan2024_rates, Mckernan2025_mcfacts}. The merging mass distribution is harder than the formation mass distribution. This is driven by the shorter inspiral time (Eq.\ref{eq:a_bin}) for more massive systems and their higher resilience to ionisations during BS and BB encounters. Both effects also lead to a flatter mass ratio distribution for mergers than formations, since it is the more massive primary BHs that form lower-mass-ratio BBHs. 

Across all three of our models, the radial distributions of formations are similar. Both our $\beta=0.5$ and $e_0=0$ models exhibit a broader radial distribution in the mergers compared to our fiducial model, driven by enhanced scattering rates, due to the larger embedded population. The maximum attainable mass is governed by the available number of objects in the disk. In our fiducial run, we find 3G+ formations/mergers are less numerous than those from lower generations, however this inverts for our $\beta=0.5$ and $e_0=0$ runs, where 3G+ mergers increase by over an order of magnitude. Again, this is driven by the greater efficiency for higher-mass BHs to pair up and merge, leading to a runaway growth of the most massive BHs in the AGN disk, once their individual masses become sufficiently disparate from the surrounding population.  

We provide a rough estimate of the rate density from our models by adopting a mean AGN number density of $n_\mathrm{AGN}\approx10^{-4}\,\mathrm{Mpc}^{-3}$, based on the SMBH accretion luminosity for our disk model $\sim 5\times 10^{43}\,\mathrm{erg}\,\mathrm{s}^{-1}$ and observations of the AGN luminosity function from \cite{Hopkins2007_AGN_n_density} (see Figure 6 therein). This gives an approximate rate of  $\mathcal{R}_\mathrm{GW}\approx2, 12, 9\,\mathrm{Gpc}^{-3}\,\mathrm{yr}^{-1}$ for our fiducial, $e_0=0$ and $\beta=0.5$ runs respectively. While this is only a crude estimate and a more realistic value would require simulating the BH population across the AGN mass function, it suggests the AGN channel can provide a non-negligible contribution to the observed rates. Considering the much flatter merger mass distribution (Figure~\ref{fig:eccentricity}), the relative contribution from the AGN channel is likely to be higher for heavier chirp mass detections.

\section{Summary and Conclusions}
We summarize the primary outcomes of our 3D $N$-body simulations as follows:
\begin{itemize}
    \item \textbf{Binary Formation and Scattering Belts:} Binary formation is highly efficient throughout the AGN disk, peaking sharply at migration traps. However, binaries at these radii rapidly undergo binary-single and binary-binary interactions. These encounters drive frequent exchanges and ionisations, scattering objects away from the trap until their orbits recircularise and realign. We term these high-intensity dynamical zones ``scattering belts''.
    
    \item \textbf{Impact of Few-Body Interactions:} Neglecting binary-single and binary-binary encounters artificially doubles the merger rate, as it permits binaries to inspiral via circumbinary disk torques without the possibility of dynamical disruption.
    
    \item \textbf{Role of Initial Eccentricity:} The initial eccentricity of the BH population critically limits merger efficiency. Modelling a realistic eccentricity distribution reduces the embedded BH population by a factor of four due to weaker accretion drag during disk transits. Consequently, both binary formation and merger rates drop by a full order of magnitude.
    
    \item \textbf{Merger Rates and Mass Distribution:} We estimate a merger rate density of $\mathcal{R}_{\mathrm{GW}} \approx 2\text{--}12\,\mathrm{Gpc}^{-3}\,\mathrm{yr}^{-1}$, highly sensitive to initial conditions. The resulting merging mass distribution is distinctly top-heavy compared to isolated and globular cluster channels, particularly when eccentricity is ignored.
    
    \item \textbf{Out-of-Disk Mergers:} A small fraction of mergers occur after the binary is dynamically ejected from the disk. These events may produce distinct electromagnetic counterparts characterized by shorter breakout and delay times compared to fully embedded mergers.
\end{itemize}

\section*{Acknowledgements}
We thank Chris Hamilton for his useful insights and comments on our approach to modelling the dynamical aspects of this problem. The research leading to this work received funding from the Independent Research Fund Denmark via grant ID 10.46540/3103-00205B. The work presented here is supported by the Carlsberg Foundation, grant CF25-1297. The Tycho supercomputer hosted at the SCIENCE HPC center at the University of Copenhagen was used for supporting this work.

\bibliography{references}
\bibliographystyle{aasjournalv7}
\end{document}